\documentclass[10pt]{article}
\usepackage[sc]{mathpazo} 
\usepackage[utf8]{inputenc}
\usepackage[T1]{fontenc}
\usepackage{cmap}
\usepackage{microtype} 
\usepackage{amssymb} 
\usepackage{amsthm} 
\usepackage{amsmath}
\usepackage[russian, english]{babel}
\usepackage[hmarginratio=1:1,top=32mm,columnsep=20pt]{geometry} 
\usepackage[hang, small,labelfont=bf,up,textfont=it,up]{caption} 
\usepackage{booktabs} 
\usepackage{float} 
\usepackage{hyperref} 
\usepackage{lettrine} 
\usepackage{paralist} 
\usepackage{abstract} 
\usepackage{titlesec} 
\titleformat{\section}[block]{\large\scshape\centering}{\thesection.}{1em}{} 
\titleformat{\subsection}[block]{\large\centering}{\thesubsection.}{1em}{} 
\numberwithin{equation}{section}

\usepackage{graphicx}
\usepackage{lipsum}
\usepackage{epstopdf}
\usepackage{comment}
\usepackage{color}
\usepackage{textcase}
\usepackage{natbib}
\usepackage{geometry}
\theoremstyle{definition}

\newtheorem*{remark*}{Remark}

\begin{document}
\begin{center}
	{\LARGE \bfseries On a new PGDUS transformed model using Inverse Weibull distribution} \\[0.5cm] 
	{\large Gauthami P. \& Chacko V. M.} \\[0.2cm] 
	{\itshape Department of Statistics \\ St. Thomas College (Autonomous), Thrissur, University of Calicut, Kerala, India - 680001} \\[0.5cm]
	\normalsize \href{mailto:gauthamistat@gmail.com}{gauthamistat@gmail.com}, 
	\normalsize \href{mailto:chackovm@gmail.com}{chackovm@gmail.com} 
\end{center}

\begin{abstract}
\noindent The Power Generalized DUS (PGDUS) Transformation is significant in reliability theory, especially for analyzing parallel systems. From the Generalized Extreme Value distribution, Inverse Weibull model particularly has wide applicability in statistics and reliability theory. In this paper we consider PGDUS transformation of Inverse Weibull distribution. The basic statistical characteristics of the new model are derived, and unknown parameters are estimated using Maximum likelihood and Maximum product of spacings methods. Simulation analysis and the reliability parameter $P(T_2<T_1)$ are explored. The effectiveness of the model in fitting a real-world dataset is demonstrated, showing better performance compared to other competing distributions.
\end{abstract}



\section{\textbf{\Large INTRODUCTION}}

To model reliability and lifetime data, a variety of distributions have been developed in the fields of statistics and reliability engineering. If the current distributions are less effective in some circumstances, researchers will inevitably have to find alternative distributions that fit the provided data. Also, it is preferable to have a suitable lifetime model when thinking about parallel systems, in which each of the components is known to have some well-known distributions. PGDUS transformation is a method for modeling parallel systems where each of the components has its distribution DUS transformed to any baseline model, which makes the transformation method more applicable in the field of reliability modeling. A model selection using DUS transformation would be more appropriate because it is a concise method for creating more appropriate distributions without adding more parameters and to fit suitable models for the given data.

Focusing on the basics of transformation techniques, sometimes, it would be better to introduce new distributions by incorporating a shape parameter, since shape of the distribution always changes dynamically from situation to situation. Let $T$ be a random variable, consider the transformation with new Cumulative Distribution Function (CDF) defined by
\begin{equation*}
    F_{T_{new}}(t)=(F_{T_{base}}(t))^\gamma
\end{equation*}
where, $F_{T_{base}}(t)$ represents the CDF of a baseline distribution and $\gamma > 0$ is an added shape parameter that enriches the model. \cite{KSS1} developed the previously mentioned DUS transformation without increasing the number of parameters and introduced the DUS Exponential (DUS-E) distribution, where this approach has proven effective in transforming baseline distributions to better reflect the complexities of real-world data. A number of studies have further explored the DUS transformation in the literature. Using both traditional and Bayesian methods, \cite{TSS1} investigated parameter estimation for the DUS-E distribution with an emphasis on upper record values. By applying the DUS transformation to the Lomax distribution, \cite{DC1} created the DUS Lomax (DUS-L) model and investigated its use in stress-strength reliability. Another significant advancement was the work of \cite{GC1}, who proposed and thoroughly examined the DUS Inverse Weibull (DUS-IW) distribution, which is distinguished by its upside-down bathtub-shaped hazard rate. Related studies that expanded these findings to the DUS-Kumaraswamy (DUS-K) distribution include those by \cite{KKKA1} and \cite{AnC1}. The applications of these transformations have been expanded by further studies by \cite{KM1}, \cite{NACAA1}, \cite{KhM1}, and \cite{HA1}, which have used them in a variety of datasets and further established their use in reliability modeling.

Modeling and Inference are fundamental in statistical research, especially in developing lifetime distributions tailored to component reliability. Many lifetime models have emerged through diverse transformation techniques, which allow for the creation of component-specific lifetime distributions. However, a more comprehensive objective lies in modeling the reliability of an entire system, especially systems with parallel-connected components. In this vein, the PGDUS transformation, introduced by \cite{TC1}, provides a valuable framework, enabling the modeling of an entire system when each component follows a DUS-transformed model. This utility of the PGDUS transformation is showcased in their proposed PGDUS-Exponential (PGDUS-E) distribution. Emphasizing this transformation, which effectively extends the DUS model through exponentiation, recent work has expanded this approach to the Weibull (PGDUS-W) and Lomax (PGDUS-L) distributions (see \citealp{TC2}) and the Inverse Kumaraswamy (PGDUS-IK) (see \citealp{AC1}) distribution.

If $T$ is the random variable under consideration, following the baseline distribution with CDF, $F_{T_{base}}(t)$ and Probability Density Function (PDF), $f_{T_{base}}(t)$, then the CDF of the PGDUS distribution is
\begin{align}
F_{T_{PGDUS}}(t)=
\begin{cases}
\left(\frac{e^{F_{T_{base}}(t)}-1}{e-1}\right)^{\gamma}
& ; \, t>0,\,\gamma>0 \\
0 & ;\, Otherwise
\end{cases}.
\label{eq:1.1}
\end{align}
The PDF of the PGDUS distribution is
\begin{align}
f_{T_{PGDUS}}(t)=
\begin{cases}
\frac{ \gamma}{ (e-1)^{\gamma}} \hspace{2mm} f_{T_{base}}(t) \, e^{F_{T_{base}}(t)} \left(e^{F_{T_{base}}(t)}-1\right)^{\gamma-1}
& ; \, t>0,\,\gamma>0 \\
0 & ;\, Otherwise
\end{cases}.
\label{eq:1.2}
\end{align}
The survival function of the PGDUS distribution is
\begin{align}
S_{T_{PGDUS}}(t)=
\begin{cases}
1-\left(\frac{e^{F_{T_{base}}(t)}-1}{e-1}\right)^{\gamma}
& ; \, t>0,\,\gamma>0 \\
0 & ;\, Otherwise
\end{cases}.
\label{eq:1.3}
\end{align}
The failure-rate function of the PGDUS distribution is
\begin{align}
h_{T_{PGDUS}}(t)=
\begin{cases}
\frac{\gamma \, f_{T_{base}}(t) \, e^{F_{T_{base}}(t)} \left(e^{F_{T_{base}}(t)}-1\right)^{\gamma-1}}{(e-1)^{\gamma}-\left(e^{F_{T_{base}}(t)}-1\right) ^{\gamma}}
& ; \, t>0,\,\gamma>0 \\
0 & ;\, Otherwise
\end{cases}.
\label{eq:1.4}
\end{align}

In this paper, we explore the PGDUS transformation with a particular emphasis on selecting an appropriate baseline model. The baseline model selection problem is quite complicated. According to the literature, the field of statistics highly values the Generalized Extreme Value (GEV) distribution family, which includes the Gumbel, Weibull, and Inverse Weibull (IW) distributions, for its versatility in reliability theory, lifetime and survival data analysis. Originally developed by Waloddi Weibull in the 1930s to characterize material strength, the Weibull distribution is still often used in statistical analysis of survival and reliability data. In their study of the three-parameter Weibull distribution, \cite{ZK1} analyzed its estimation procedures and pointed out that it can be difficult to estimate all three parameters at once, which frequently causes researchers to rely on the more straight-forward two-parameter Weibull distribution. Similarly, constraints have been identified in the Maximum Likelihood (ML) estimation for the three-parameter Weibull model in early researches (see \citealp{H1,H2}). Due in large part to its adaptability in modeling a variety of experimental data, subsequent research, including as studies by \cite{MSK1} and \cite{EKG1}, has made a substantial contribution to the understanding of the Weibull distribution. This distribution transforms into the IW distribution, sometimes referred to as the Fréchet (F) distribution, when inverse transformation is applied, and is characterized by its three-parameter CDF, \eqref{eq:1.5} and PDF, \eqref{eq:1.6}.
\begin{align}
F_{T_{F}}(t)=
\begin{cases}
e^{-\left(\frac{t-\alpha}{\theta}\right)^{-\lambda}}
& ; \, t>0,\,\lambda>0,\, \theta>0, \alpha \in \mathbb{R} \\
0 & ;\, Otherwise
\end{cases}.
\label{eq:1.5}
\end{align}

\begin{align}
f_{T_{F}}(t)=
\begin{cases}
\frac{\lambda}{\theta}\, \left(\frac{t-\alpha}{\theta}\right)^{-(\lambda+1)}\, e^{-\left(\frac{t-\alpha}{\theta}\right)^{-\lambda}}
& ; \, t>0,\,\lambda>0,\, \theta>0, \alpha \in \mathbb{R} \\
0 & ;\, Otherwise
\end{cases}.
\label{eq:1.6}
\end{align}
This is simply a two-parameter IW model when the third parameter $\alpha$ equals zero, which considered to be our baseline distribution.

With regard to its flexibility in expressing data with monotonic failure rates, the IW distribution is a flexible lifetime model that is widely used in domains such as engineering, survival analysis, reliability theory, and risk modeling. The IW model has proven useful for practical applications, much like the Weibull distribution. \cite{KKP1}, \cite{CP2} and \cite{CP1} are the early research works on IW distribution. \cite{MSA1} examined order statistics and related inference for the IW model, which is well-documented in statistical literature and goes by several names, such as complementary Weibull, reciprocal Weibull, and reverse Weibull (see \citealp{D1}, \citealp{MK1}, and \citealp{MBE1}). Studies by \cite{KPP1} and \cite{dGOC1} further emphasize the many uses of the IW distribution. Recent studies include \cite{ASA1}, who employed the IW model for fitting wind speed data as a replacement to the Weibull distribution; \cite{OETB1}, who proposed the Marshall–Olkin extended IW model; \cite{LU1}, who compared methods for IW parameter estimation; \cite{EY1}, who looked at new properties and applications of the IW distribution; and \cite{RLRD1}, who gave an overview of the estimation and applications of the Fréchet distribution.

Given the wide applicability of both the PGDUS transformation technique and the IW distribution, we introduce a new model: the $PGDUS-IW (\lambda, \theta, \gamma) $ distribution in Section~\ref{Sec2}, where $\theta$ denotes the scale parameter, and $\lambda$ and $\gamma$ represent the shape parameters. The fundamental statistical properties of this new model have been derived and examined, including plotting the PDF and failure-rate functions to illustrate its behavior. Our next goal is to estimate the model parameters, for which we apply ML estimation. ML estimation, among the most reliable techniques, is providing consistent, asymptotically normal, and asymptotically efficient estimators for large samples under general conditions. The baseline model in our study is the two-parameter IW distribution (scale-shape), chosen for its simplicity and efficiency in ML parameter estimation, as models without location parameters can often lead to better performance in this context and since alternative methods can address potential challenges in parameter estimation, we also explore the Maximum Product of Spacings (MPS) method, detailing estimation procedures, all given in Section~\ref{Sec3} and conducting simulation studies given in Section~\ref{Sec4} to compare these approaches. Furthermore, a real-data investigation was carried out to evaluate the performance of the PGDUS-IW model in comparison to other PGDUS-based models with three parameters, demonstrating its efficacy across several real-world datasets. The probability, $P(T_2 < T_1)$, in single and multi-component models are further investigated and is given in Section~\ref{Sec5} and the estimation using ML and MPS techniques, and the results are confirmed through simulation and real-data analysis. Moreover, the main conclusions are given in Section~\ref{Sec6}.

\section{\textbf{\Large $PGDUS-IW (\lambda, \theta, \gamma) $ DISTRIBUTION}}
\label{Sec2}

In this section, we propose the new model, $PGDUS-IW (\lambda, \theta, \gamma)$, where the baseline model has the CDF and PDF, respectively,
\begin{align}
F_{T_{IW}}(t)=
\begin{cases}
e^{-\left(\frac{t}{\theta}\right)^{-\lambda}}
& ; \, t>0,\,\lambda>0,\, \theta>0 \\
0 & ;\, Otherwise
\end{cases}
\label{eq:2.1}
\end{align}
and
\begin{align}
f_{T_{IW}}(t)=
\begin{cases}
\frac{\lambda}{\theta}\, \left(\frac{t}{\theta}\right)^{-(\lambda+1)}\, e^{-\left(\frac{t}{\theta}\right)^{-\lambda}}
& ; \, t>0,\,\lambda>0,\, \theta>0\\
0 & ;\, Otherwise
\end{cases}.
\label{eq:2.2}
\end{align}

\noindent Using \eqref{eq:1.1}, \eqref{eq:1.2},\eqref{eq:1.3},\eqref{eq:1.4}, \eqref{eq:2.1} and \eqref{eq:2.2}, we get the new model's respective functions as given below.
\begin{align}
F_{T_{PGDUS-IW}}(t)=
\begin{cases}
\left(\frac{e^{e^{-\left(\frac{t}{\theta}\right)^{-\lambda}}}-1}{e-1}\right)^{\gamma}
& ; \, t>0,\,\lambda,\,\theta,\,\gamma>0 \\
0 & ;\, Otherwise
\end{cases},
\label{eq:2.3}
\end{align}

\begin{align}
f_{T_{PGDUS-IW}}(t)=
\begin{cases}
\frac{ \gamma \lambda \theta^\lambda}{ (e-1)^{\gamma}} \hspace{2mm} t^{-(\lambda+1)} \hspace{2mm} e^{-\left(\frac{t}{\theta}\right)^{-\lambda}} \hspace{2mm}e^{e^{-\left(\frac{t}{\theta}\right)^{-\lambda}}} \left(e^{e^{-\left(\frac{t}{\theta}\right)^{-\lambda}}}-1\right)^{\gamma-1}
& ; \, t>0,\,\lambda,\,\theta,\,\gamma>0 \\
0 & ;\, Otherwise
\end{cases},
\label{eq:2.4}
\end{align}

\begin{align}
S_{T_{PGDUS-IW}}(t)=
\begin{cases}
1-\left(\frac{e^{e^{-\left(\frac{t}{\theta}\right)^{-\lambda}}}-1}{e-1}\right)^{\gamma}
& ; \, t>0,\,\lambda,\,\theta,\,\gamma>0 \\
0 & ;\, Otherwise
\end{cases},
\label{eq:2.5}
\end{align}
and
\begin{align}
h_{T_{PGDUS-IW}}(t)=
\begin{cases}
\frac{\gamma \lambda \theta^\lambda \hspace{2mm} t^{-(\lambda+1)} e^{-\left(\frac{t}{\theta}\right)^{-\lambda}}\hspace{2mm}e^{e^{-\left(\frac{t}{\theta}\right)^{-\lambda}}} \left(e^{e^{-\left(\frac{t}{\theta}\right)^{-\lambda}}}-1\right)^{\gamma-1}}{(e-1)^{\gamma}-\left(e^{e^{-\left(\frac{t}{\theta}\right)^{-\lambda}}}-1\right) ^{\gamma}}
& ; \, t>0,\,\lambda,\,\theta,\,\gamma>0 \\
0 & ;\, Otherwise
\end{cases}.
\label{eq:2.6}
\end{align}

The PDF and failure rate plots are plotted in such a way that all the combinations of parameter values are considered such as $\theta \leq \gamma \leq \lambda$, $\gamma \leq \theta \leq \lambda$, $\lambda \leq \gamma \leq \theta$, $\lambda \leq \theta \leq \gamma  $, $  \gamma \leq \lambda \leq \theta$ and $\theta  \leq \lambda \leq \gamma$ and are given in Figure~\ref{Fig1} and Figure~\ref{Fig2} respectively, wherein l denote $\lambda$, t denote $\theta$ and g denote $\gamma$. From Figure~\ref{Fig2}, we can see that the failure-rate function has both upside-down and monotonically decreasing nature, which is reliable in lifetime modeling.
\newpage
\begin{figure}[h]
	\centering
	\caption{PDF Plot of PGDUS-IW ($\lambda$, $\theta$, $\gamma$) distribution}
	\label{Fig1}
\end{figure}

\begin{figure}[h]
	\centering
	\caption{Failure-Rate plot of PGDUS-IW ($\lambda$, $\theta$, $\gamma$) distribution}
	\label{Fig2}
\end{figure}

In addition to the basic functions of the model, several key statistical properties and information measures have been derived with usual notations. These include:
\begin{itemize}
	\item Quantile function: The $p^{th}$ quantile is,
	\begin{equation*}
	q_p=\frac{(-1)^{-\frac{1}{\lambda}} \theta}{\left\{\ln \left[\ln \left(1+(e-1) p^{\frac{1}{\gamma}}\right) \right] \right\}^{\frac{1}{\lambda}}} \, ; \, 0<p<1.
	\end{equation*}
	\item Moment Generating Function (MGF):
	\begin{equation*}
	M_{T_{PGDUS-IW}}(s)=\frac{\gamma}
	{ (e-1)^{\gamma}} \,\sum_{k=0}^{\gamma-1} \sum_{m=0}^{\infty}\sum_{n=0}^{\infty}  \,(-1)^{k-1} \binom{\gamma-1}{k}\,\frac{(\gamma-k)^{m} s^{n} \theta^{n}}{m!n!}\frac{\Gamma\left(-\frac{n}{\lambda}+1\right)}{(m+1)^{-\frac{n}{\lambda}+1}} \, ;\, n<\lambda.
	\end{equation*}
	\item Characteristic Function (CF):
	\begin{equation*}
	\Phi_{T_{PGDUS-IW}}(s)=\frac{\gamma}
	{ (e-1)^{\gamma}} \,\sum_{k=0}^{\gamma-1} \sum_{m=0}^{\infty}\sum_{n=0}^{\infty}  \,(-1)^{k-1} \binom{\gamma-1}{k}\,\frac{(\gamma-k)^{m} (is)^{n} \theta^{n}}{m!n!}\frac{\Gamma\left(-\frac{n}{\lambda}+1\right)}{(m+1)^{-\frac{n}{\lambda}+1}} \, ;\, n<\lambda,
	\end{equation*}
	where $i=\sqrt{-1}$.
	\item Cumulant Generating Function (CGF):
	\begin{eqnarray}
	K_{T_{PGDUS-IW}}(s)&=&\ln \gamma-\gamma \ln(e-1) \nonumber \\
	&& +\ln\left[\sum_{k=0}^{\gamma-1} \sum_{m=0}^{\infty}\sum_{n=0}^{\infty}  \,(-1)^{k-1} \binom{\gamma-1}{k}\,\frac{(\gamma-k)^{m} s^{n} \theta^{n}}{m!n!}\frac{\Gamma\left(-\frac{n}{\lambda}+1\right)}{(m+1)^{-\frac{n}{\lambda}}+1}\right] \,;\, n<\lambda. \nonumber
	\end{eqnarray}
	\item Raw moments: The $s^{th}$ raw moment is,
	\begin{equation*}
	\mu_{s}^{\prime}=\frac{\gamma \theta^{s} }{(e-1)^{\gamma}}\,\sum_{k=0}^{\gamma-1}\sum_{m=0}^{\infty} \,(-1)^{k-1} \binom{\gamma-1}{k}\,\frac{(\gamma-k)^{m} }{m!}\frac{\Gamma\left(-\frac{s}{\lambda}+1\right)}{(m+1)^{-\frac{s}{\lambda}+1}} \, ; \, s<\lambda.
	\end{equation*}
	\item Information measures:
	\begin{itemize}
		\item Rényi entropy of order $\delta$:
		\begin{eqnarray}
		H_{\delta}(T)&=& \frac{1}{1-\delta} \ln \left[ \int_{t}^{} f^\delta (t) \, dt \right] \nonumber \\
		&=&\frac{\delta}{1-\delta} \ln \gamma - \ln \lambda + \ln \theta - \frac{\delta \gamma}{1-\delta} \ln (e-1) \nonumber \\  && + \frac{1}{1-\delta} \, \ln \left[ \sum_{k=0}^{\delta(\gamma-1)}\sum_{m=0}^{\infty} \, \frac{(-1)^{k-1}}{m!} \binom{\delta(\gamma-1)}{k}\,(\delta\gamma-k)^{m} \frac{\Gamma\left(\delta+\frac{\delta-1}{\lambda}\right)}{(m+\delta)^{\delta+\frac{\delta-1}{\lambda}}} \right], \nonumber
		\end{eqnarray}
		where $\delta>0$ and $ \delta \neq 1 $.
		\item Extropy (Compliment dual of Shannon Entropy, see \citealp{LSA1}):
		\begin{eqnarray}
		J(T)&=& -\frac{1}{2}
		\int_{t}^{} f^2 (t) \, dt \nonumber \\
		&=& \frac{-(\lambda+1) \gamma^2}{2 \theta (e-1)^{2\gamma}} \sum_{k=0}^{2(\gamma-1)}\sum_{m=0}^{\infty} \, \frac{(-1)^{k-1}}{m!} \binom{2(\gamma-1)}{k}\,(2\gamma-k)^{m} \frac{\Gamma\left(1+\frac{1}{\lambda}\right)}{(m+2)^{2+\frac{1}{\lambda}}}. \nonumber
		\end{eqnarray}
	\end{itemize}
\end{itemize}

\subsection{Distribution of Order Statistic}
Taking randomly a sample of size $n$ from $PGDUS-IW (\lambda, \theta, \gamma) $ distribution, the $r^{th}$ order statistic, where, $r=1,2,...,n$ will have the CDF and PDF as given in \eqref{eq:2.7} and \eqref{eq:2.8} respectively.
\begin{eqnarray}
F_{T_{PGDUS-IW_{r:n}}}(t)&=&
\frac{1}{(e-1)^{n\gamma}} \sum_{k=r}^{n} \binom{n}{k} \left[e^{e^{-\left(\frac{t}{\theta}\right)^{-\lambda}}}-1\right]^{\gamma k} \label{eq:2.7} \\ && \times
\left[(e-1)^{\gamma}-\left(e^{e^{-\left(\frac{t}{\theta}\right)^{-\lambda}}}-1\right)^{\gamma} \right]^{n-k}  ;  t>0, \, \lambda,\,\theta,\,\gamma>0. \nonumber
\end{eqnarray}
\begin{align}
f_{T_{PGDUS-IW_{r:n}}}(t)=
\begin{cases}
    \frac{n!}{(r-1)!(n-r)!} \frac{ \gamma \lambda \theta^\lambda}{ (e-1)^{n \gamma}} \hspace{2mm} t^{-(\lambda+1)} \hspace{2mm} e^{-\left(\frac{t}{\theta}\right)^{-\lambda}} \hspace{2mm}e^{e^{-\left(\frac{t}{\theta}\right)^{-\lambda}}} \vspace{1mm} \\ \left[e^{e^{-\left(\frac{t}{\theta}\right)^{-\lambda}}}-1\right]^{\gamma r -1} \left[(e-1)^{\gamma}-\left(e^{e^{-\left(\frac{t}{\theta}\right)^{-\lambda}}}-1\right)^{\gamma} \right]^{n-r} & ;\, t>0, \, \lambda,\,\theta,\,\gamma>0 \vspace{2mm} \\
    0 & ;\, Otherwise
\end{cases}.
\label{eq:2.8}
\vspace{3mm}
\end{align}

\section{\textbf{ESTIMATION OF PARAMETERS}}
\label{Sec3}
The unknown parameters of the $PGDUS-IW (\lambda, \theta, \gamma)$ univariate distribution have to be estimated. The two equivalent estimation techniques that are commonly employed in practice are ML estimation and MPS estimation. ML estimation seeks to identify the parameter values that maximize the likelihood function, which represents the probability of observing the given data under a specific set of parameters. The parameter estimates obtained by ML estimation are considered to be asymptotically unbiased, consistent, and efficient under regular conditions.

MPS estimation is an alternative to ML estimation and is based on the concept of spacings, which are the gaps between ordered data points, along with the CDF. That is, ML estimation uses the raw density values, whereas MPS estimation uses the cumulative properties to consider the relative spacing between data points. If ML method finds out the statistic that maximizes the likelihood function, MPS method also finds out the statistic that maximize the product of the spacings between ordered data points, offering robustness in more challenging scenarios. In terms of the maximization process, both approaches are equivalent.
\subsection{ML estimation procedure}
Observe a random sample of size $n$, $T_1, T_2, . . . , T_n$,  from a $PGDUS-IW (\lambda, \theta, \gamma)$ distribution where the likelihood function is
\begin{equation*}
    L(\lambda, \theta, \gamma \mid t_{i})=\prod_{i=1}^{n} f_{T_{PGDUS-IW}}(t_{i}, \lambda, \theta, \gamma).
\end{equation*}
On substitution, we have the likelihood function,
\begin{align}
    L=\frac{\gamma^n \lambda^n \theta^{\lambda n}}{(e-1)^{\gamma n}} \, \left[\prod_{i=1}^{n} t_i \right]^{-(\lambda+1)} \, e^{-\sum_{i=1}^{n}\left(\frac{t_{i}}{\theta}\right)^{-\lambda}}\, e^{\sum_{i=1}^{n} e^{-\left(\frac{t_{i}}{\theta}\right)^{-\lambda}}} \left[\prod_{i=1}^{n} \left(e^{e^{-\left(\frac{t_{i}}{\theta}\right)^{-\lambda}}}-1 \right)\right]^{\gamma-1}.
    \label{eq:3.1}
\end{align}
Maximization of the likelihood function is similar to the maximization of log-likelihood (log-L) function and the partial derivatives to be solved for finding the ML estimator(MLE)s of each of the parameters are found out and is tedious as those are not in the explicit form but can be solved using numerical techniques, but for the parameter $\gamma$, we get,
\begin{equation}
    \hat{\gamma}^{MLE}=\frac{n}{n \ln (e-1)-\sum_{i=1}^{n}\ln\left(e^{e^{-\left(\frac{t_{i}}{\theta}\right)^{-\lambda}}}-1\right)},
    \label{eq:3.2}
\end{equation}
which is the MLE of $\gamma$.

\subsection{MPS estimation procedure}
The method of MPS estimation suggested in \cite{CA1} and following the procedure, an ordered random sample with size $n$, $T_{1:n} < T_{2:n} < . . . < T_{n:n}$, taken from $PGDUS-IW (\lambda, \theta, \gamma)$ distribution. The spacings are calculated using the formula,
\begin{align}
    S_i=F_{T_{PGDUS-IW}} \left(t_{i:n}\right)-F_{T_{PGDUS-IW}} \left(t_{i-1:n}\right);\, i=1,2,...,n+1,
    \label{eq:3.3}
\end{align}
where $F_{T_{PGDUS-IW}}(t_{0:n})=0$ and $F_{T_{PGDUS-IW}}(t_{n+1:n})=1$.  \\
Therefore,
\begin{eqnarray}
    S_1&=&F_{T_{PGDUS-IW}} \left(t_{1:n}\right)-F_{T_{PGDUS-IW}} \left(t_{0:n}\right) \nonumber \\ &=& F_{T_{PGDUS-IW}} \left(t_{1:n}\right) \nonumber \\
    S_i&=&F_{T_{PGDUS-IW}} \left(t_{i:n}\right)-F_{T_{PGDUS-IW}} \left(t_{i-1:n}\right);\, i=2,...,n \nonumber\\
    S_{n+1}&=&F_{T_{PGDUS-IW}} \left(t_{n+1:n}\right)-F_{T_{PGDUS-IW}} \left(t_{n:n}\right) \nonumber \\ &=& 1 - F_{T_{PGDUS-IW}} \left(t_{n:n}\right) \nonumber
\end{eqnarray}
MPS estimator(MPSE)s are those which maximizes the geometric mean of the spacings. i.e., we need to maximize the product of spacings (PS) function,
\begin{equation*}
    PS(\lambda, \theta, \gamma) = \left[ \prod_{i=1}^{n+1} S_i \right]^{\frac{1}{n+1}}
\end{equation*}
\noindent or, the log-PS function,
\begin{equation*}
    \ln PS(\lambda, \theta, \gamma)= \frac{1}{n+1} \sum_{i=1}^{n+1} \ln(S_i)
\end{equation*}
which can be re-written as
\begin{multline}
    \ln PS(\lambda, \theta, \gamma)= (n+1)^{-1} \Biggl\{\ln F_{T_{PGDUS-IW}} \left(t_{1:n}\right)+ \ln\left( 1 - F_{T_{PGDUS-IW}} \left(t_{n:n}\right) \right) \\ + \sum_{i=2}^{n} \ln \left( F_{T_{PGDUS-IW}} \left(t_{i:n}\right)-F_{T_{PGDUS-IW}} \left(t_{i-1:n}\right)  \right)
     \Biggl\}.
    \label{eq:3.4}
\end{multline}
The $PGDUS-IW (\lambda, \theta, \gamma)$ distribution has the log-PS function
\begin{eqnarray}
    \ln PS(\lambda, \theta, \gamma)&=& (n+1)^{-1} \Biggl\{ \gamma \ln \left( e^{e^{-\left(\frac{t_{1:n}}{\theta}\right)^{-\lambda}}}-1 \right) + \ln\left[ (e-1)^{\gamma} - \left( e^{e^{-\left(\frac{t_{n:n}}{\theta}\right)^{-\lambda}}}-1 \right)^{\gamma} \right]  \nonumber \\
    && + \sum_{i=2}^{n} \ln \left[ \left( e^{e^{-\left(\frac{t_{i:n}}{\theta}\right)^{-\lambda}}}-1 \right)^{\gamma} - \left( e^{e^{-\left(\frac{t_{i-1:n}}{\theta}\right)^{-\lambda}}}-1 \right)^{\gamma} \right] - (n+1)\, \gamma \,\ln(e-1)\Biggl\}. \nonumber \\ \label{eq:3.5}
\end{eqnarray}
\noindent Partial differentiation of \eqref{eq:3.5} with respect to each of the parameters and equating them to zero will yield the corresponding MPSEs, but cannot be solved explicitly.

\section{\textbf{SIMULATION STUDY \& DATA ANALYSIS}}
\label{Sec4}
The estimation techniques studied do not explicitly produce the estimators of the unknown parameters. Using an initialization for the parameters values, we carry out a simulation study using both ML and MPS methods, and to assess the performance of the model, we simulate 1000 samples of sizes 50, 100, 150, and 350 from the $PGDUS-IW (\lambda, \theta, \gamma)$ model, using the  result that every distribution function follows a $u(0, 1)$ model. The bias and mean square error (MSE) of the parameter estimates are computed, specifically, choosing the parameter combination (1, 0.6, 0.3). The results are presented in Table~\ref{Tab1}.
\begin{table}[h]
        \centering
        \begin{tabular}{cccc}
        \hline
        \multicolumn{4}{c}{ML Method}  \\
        \hline
          n   & $\hat{\lambda}$ & bias ($\hat{\lambda}$) & MSE ($\hat{\lambda}$) \\
          \hline
          50 & $1.0246$  & $0.0246$       & $0.0141$   \\
          100 & $1.0073$   &$0.0073$ & $0.0062$ \\
          150 & $1.0034$  &$0.0034$ & $0.0041$ \\
          350 & $0.9989$ &$-0.0011$ &  $0.0017$\\
          \hline
          n   & $\hat{\theta}$ & bias ($\hat{\theta}$) & MSE ($\hat{\theta}$) \\
          \hline
          50 &  $0.7107$   &$0.1107$ &$0.3281$      \\
          100 & $0.6801$ & $0.0801$&$0.1867$ \\
          150 & $0.6331$&$0.0331$ &$0.1308$  \\
          350 & $0.5923$ & $-0.0077$& $0.0717$ \\
          \hline
          n   & $\hat{\gamma}$ & bias ($\hat{\gamma}$) & MSE ($\hat{\gamma}$) \\
          \hline
          50 & $0.6515$ &$0.3515$ &  $2.3226$ \\
          100 & $0.5372$&$0.2372$ & $1.1418$ \\
          150 &$0.4968$ & $0.1968$ & $0.5665$\\
          350 & $0.4711$& $0.1711$  & $0.5408$\\
          \hline
        \multicolumn{4}{c}{MPS Method} \\
        \hline
          n   & $\hat{\lambda}$ & bias ($\hat{\lambda}$) & MSE ($\hat{\lambda}$) \\
          \hline
          50 &   $0.9511$    & $-0.0489$    &  $0.0144$    \\
          100 &  $0.9619$  &$-0.0380$ & $0.0072$ \\
          150 &  $0.9699$ &  $-0.0301$ & $0.0049$ \\
          350 & $0.9816$&  $-0.0184$ &  $0.0022$ \\
          \hline
          n   & $\hat{\theta}$ & bias ($\hat{\theta}$) & MSE ($\hat{\theta}$) \\
          \hline
          50 & $0.8376$ &  $0.2376$   &   $0.7693$   \\
          100 & $0.7164$ & $0.1164$& $0.3414$ \\
          150 &$0.6339$ &$0.0339$&  $0.2134$\\
          350 & $0.5874$ & $-0.0126$& $0.1062$  \\
          \hline
          n   & $\hat{\gamma}$ & bias ($\hat{\gamma}$) & MSE ($\hat{\gamma}$) \\
          \hline
          50 & $0.9850$ &  $0.6850$   &  $4.3225$ \\
          100 &$0.8876$&$0.5876$ &   $3.0054$ \\
          150 &   $0.8083$& $0.5083$&  $1.9639$ \\
          350 & $0.6673$ & $0.3673$&$1.3958$ \\
          \hline
        \end{tabular}
        \caption{Results of simulation study using both ML and MPS methods}
        \label{Tab1}
\end{table}

From the Table~\ref{Tab1}, we can see that as the sample size increases the bias and MSE calculated using both ML and MPS methods decreases for each of the parameter estimates. And the Table~\ref{Tab1} accurately shows that the ML estimates have better values than MPS estimates for the proposed model.

From the real-world scenarios, we consider a dataset for fitting using $PGDUS-IW (\lambda, \theta, \gamma)$ model and compare the results with those obtained by fitting the same data using the already proposed PGDUS-based models with three parameters such as PGDUS-IK, PGDUS-W and PGDUS-L distributions. The dataset considered is taken from \cite{GrC1} which is also used in \cite{AAAOH1}. Table~\ref{Tab2} given below explicitly shows the dataset.
\begin{table}[h]
  \centering
  \begin{tabular}{cccccccccc}
  \hline
  1.1 & 1.4 & 1.3 & 1.7 & 1.9 & 1.8 & 1.6 &  2.2 &  1.7 & 2.7\\
  4.1 & 1.8 & 1.5 & 1.2 & 1.4 & 3 & 1.7 & 2.3 & 1.6 & 2\\
  \hline
\end{tabular}
\caption{Relief times of 20 patients taken for data analysis}
\label{Tab2}
\end{table}

Using R programming language, the analysis is carried out and the results are given in Table~\ref{Tab3} and Table~\ref{Tab4}, where we have carried out three goodness of fit tests, Kolmogorov-Smirnov (KS) test, Anderson-Darling (AD) test, and Cramer-von Mises (CVM) test and computed the criteria for small samples, corrected Akaike information criterion (AICc) and corrected Bayesian
information criterion (BICc). The perfection of the competing models is tested, conclusion obtained and the results shows that $PGDUS-IW (\lambda, \theta, \gamma)$ distribution has the highest p-value on all the tests carried out and the least AICc and BICc values comparing to the competing models in both ML and MPS methods. Moreover, the ML method yields better results and hence its e-CDF, $F_{PGDUS-IW_{n}}(t)$, and fitted PDF plots are given in Figure~\ref{Fig3} and Figure~\ref{Fig4} respectively for visual comparison.

\newpage
\begin{figure}[h]
    \centering
    \caption{Theoretical and Empirical CDFs}
    \label{Fig3}
\end{figure}

\begin{figure}[h]
    \centering
    \caption{Fitted Densities}
    \label{Fig4}
\end{figure}

\section{\textbf{COMPUTATION AND INFERENCE OF THE PARAMETER $P(T_2<T_1)$}}
\label{Sec5}
We consider computing the reliability parameter, $P(T_2<T_1)$, in both single component and multi-component models. In particular, from the literature, we can term $T_1$ as `Strength' and $T_2$ as `Stress' components, but here we focus on those variables as such due to the limitation of availability on Stress-Strength components' real-world datasets. \cite{CH1} and \cite{BJ1} looked over the early research on Stress-Strength reliability and how to estimate it. This idea was covered by Kundu and Gupta in their work in 2005 on the generalized exponential distribution (\citealp{KG1}) and their work in 2006 on the Weibull distribution (\citealp{KG2}). While \cite{RMK1} relied on the three-parameter Generalized Exponential distribution for the stress-strength parameter estimation, \cite{KR1} calculated this reliability parameter and estimated the same for the three-parameter Weibull distribution. \cite{G1} discusses the Generalized Exponential distribution's multi-component parameter. \cite{AGK1}, \cite{SD1}, \cite{AAN1}, \cite{AA1}, \cite{AAAAMA1}, \cite{JDAAT1}, \cite{AAR1}, \cite{S1}, \cite{CE1}, and \cite{SCK1} have all discussed the computation, estimation, and application procedures for this reliability parameter for a variety of existing distributions over the past ten years. In many of these studies, different estimation techniques, including ML, MPS, and Bayesian, as well as data analysis, including different censoring schemes, have been covered. Additionally, in between, \cite{PA1} computed the relevant reliability parameter for parallel systems with independent and non-identical components, and \cite{SK1} developed conditional stress-strength reliability concept.
\subsection{Single Component, $R$}
Let $T_1 \sim PGDUS-IW (\lambda, \theta, \gamma_1)$ distribution and $T_2 \sim PGDUS-IW (\lambda, \theta, \gamma_2)$ distribution, chosen as such, for the ease of calculations.
Then, the single component system reliability $R$ is
\begin{eqnarray}
  R &=& P(T_2<T_1)\nonumber \\
  &=& \int_{t}^{} f_{T_1}(t) \, F_{T_2}(t) \cdot dt \nonumber \\
  &=& \frac{\gamma_1 \lambda \theta^\lambda}{(e-1)^{\gamma_1 + \gamma_2}} \int_{0}^{\infty} t^{-(\lambda+1)}\, e^{-\left( \frac{t}{\theta} \right)^{-\lambda}} e^{e^{-\left( \frac{t}{\theta} \right)^{-\lambda}}} \left[e^{e^{-\left( \frac{t}{\theta} \right)^{-\lambda}}} -1 \right]^{\gamma_1 +\gamma_2 -1} \cdot dt \nonumber
\end{eqnarray}
After further substitution and simplification, we get,
\begin{equation}\label{eq:5.1}
         R= \frac{\gamma_1}{\gamma_1 + \gamma_2}; \, \gamma_1,\, \gamma_2 >0.
\end{equation}
\subsubsection{ML Estimation of $R$}
We take a random sample of size $N$, $T_{1_1}, T_{1_2}, ..., T_{1_N}$, from $PGDUS-IW(\lambda, \theta, \gamma_1)$ and a random sample of size $M$, $T_{2_1}, T_{2_2}, ..., T_{2_M}$, from $PGDUS-IW(\lambda, \theta, \gamma_2)$ distributions. To estimate $R$, we need to find the estimators of $\gamma_1$ and $\gamma_2$. Here, the likelihood function is
\begin{equation*}
    L(\lambda, \theta, \gamma_1, \gamma_2 \mid t_{1_i}, t_{2_j})=\prod_{i=1}^{N} f_{T_{1_{PGDUS}}}(t_{1_i}, \lambda, \theta, \gamma_1) \, \prod_{j=1}^{M} f_{T_{2_{PGDUS}}}(t_{2_j}, \lambda, \theta, \gamma_2).
\end{equation*}
On required substitution, simplification and taking logarithm, we get the log-L function,

\begin{eqnarray}
    \ln L&=& N \ln \gamma_1 + M \ln \gamma_2 + (N+M) (\ln \lambda + \lambda \ln \theta) - (N \gamma_1 + M \gamma_2) \ln (e-1) \nonumber \\
    && - (\lambda+1) \left[ \sum_{i=1}^{N} \ln t_{1_i} + \sum_{j=1}^{M} \ln t_{2_j}\right] - \sum_{i=1}^{N} \left(\frac{t_{1_i}}{\theta}\right)^{-\lambda}- \sum_{j=1}^{M} \left(\frac{t_{2_j}}{\theta}\right)^{-\lambda} + \sum_{i=1}^{N} e^{-\left(\frac{t_{1_i}}{\theta}\right)^{-\lambda}} \nonumber  \\
    &&  + \sum_{j=1}^{M} e^{-\left(\frac{t_{2_j}}{\theta}\right)^{-\lambda}} + (\gamma_1-1)\sum_{i=1}^{N} \ln \left(e^{e^{-\left(\frac{t_{1_i}}{\theta}\right)^{-\lambda}}} -1\right) + (\gamma_2-1)\sum_{j=1}^{M} \ln \left(e^{e^{-\left(\frac{t_{2_j}}{\theta}\right)^{-\lambda}}} -1\right). \nonumber \\ \label{eq:5.2}
\end{eqnarray}
\noindent Equating the partial derivatives of log-L with respect to the unknown parameters to be estimated, and equating them to zero will yield the corresponding MLEs,
\begin{equation}
    \hat{\gamma_1}^{MLE}=\frac{N}{N \ln (e-1)-\sum_{i=1}^{N}\ln \left(e^{e^{-\left(\frac{t_{1_i}}{\theta}\right)^{-\lambda}}} -1\right)},
    \label{eq:5.3}
\end{equation}
which is the MLE of $\gamma_1$ and
\begin{equation}
    \hat{\gamma_2}^{MLE}=\frac{M}{M \ln (e-1)-\sum_{j=1}^{M}\ln \left(e^{e^{-\left(\frac{t_{2_j}}{\theta}\right)^{-\lambda}}} -1\right)},
    \label{eq:5.4}
\end{equation}
which is the MLE of $\gamma_2$. Hence, the MLE of $R$ is
\begin{equation}
    \hat{R}^{MLE}=\frac{\hat{\gamma_1}^{MLE}}{\hat{\gamma_1}^{MLE} + \hat{\gamma_2}^{MLE}}.
    \label{eq:5.5}
\end{equation}

\subsubsection{MPS Estimation of $R$}
An ordered random sample $T_{1_{1:N}} < T_{1_{2:N}} < . . . < T_{1_{N:N}}$ taken from $PGDUS-IW (\lambda, \theta, \gamma_1)$ distribution and another one $T_{2_{1:M}} < T_{2_{2:M}} < . . . < T_{2_{M:M}}$ from $PGDUS-IW (\lambda, \theta, \gamma_2)$ distribution.
The spacings of $T_1$'s distribution:
\begin{eqnarray}
    S_{1}^{1}&=&\left(\frac{e^{e^{-\left(\frac{t_{1_{1:N}}}{\theta}\right)^{-\lambda}}}-1}{e-1}\right)^{\gamma_1} \nonumber \\
    S_{i}^{1}&=&\left(\frac{e^{e^{-\left(\frac{t_{1_{i:N}}}{\theta}\right)^{-\lambda}}}-1}{e-1}\right)^{\gamma_1}-\left(\frac{e^{e^{-\left(\frac{t_{1_{i-1:N}}}{\theta}\right)^{-\lambda}}}-1}{e-1}\right)^{\gamma_1}
    \,;\, i=2,...,N \nonumber \\
    S_{N+1}^{1}&=& 1 - \left(\frac{e^{e^{-\left(\frac{t_{1_{N:N}}}{\theta}\right)^{-\lambda}}}-1}{e-1}\right)^{\gamma_1} \nonumber
\end{eqnarray}
These spacings are used to obtain the PS function ($PS_1$). Taking logarithm and on simplifying, the corresponding log-PS function becomes
\begin{eqnarray}
    \ln PS_1(\lambda, \theta, \gamma_1) &=&  (N+1)^{-1} \Biggl\{ \gamma_1 \ln \left( e^{e^{-\left(\frac{t_{1_{1:N}}}{\theta}\right)^{-\lambda}}}-1 \right) \nonumber \\
    && + \sum_{i=2}^{N} \ln \left[ \left( e^{e^{-\left(\frac{t_{1_{i:N}}}{\theta}\right)^{-\lambda}}}-1 \right)^{\gamma_1} - \left( e^{e^{-\left(\frac{t_{1_{i-1:N}}}{\theta}\right)^{-\lambda}}}-1 \right)^{\gamma_1} \right] \label{eq:5.6}  \\
    &&  + \ln\left[ (e-1)^{\gamma_1} - \left( e^{e^{-\left(\frac{t_{1_{N:N}}}{\theta}\right)^{-\lambda}}}-1 \right)^{\gamma_1} \right] - (N+1)\, \gamma_1 \,\ln(e-1)\Biggl\}. \nonumber
\end{eqnarray}
In the similar way (with similar notations), the spacings of $T_2$'s distribution ($S_{j}^{2}, j=1, 2, ..., M$) and corresponding PS function ($PS_2$) can be derived out.
The log-PS function of the bivariate model to be maximized is
\begin{equation*}
    \ln PS(\lambda, \theta, \gamma_1, \gamma_2) = \ln PS_1(\lambda, \theta, \gamma_1) + \ln PS_2(\lambda, \theta, \gamma_2).
\end{equation*}
Therefore, the function to be maximized to obtain the estimators of $\gamma_1$, $\gamma_2$ and so the reliability parameter $R$ is
\begin{eqnarray}
    \ln PS(\lambda, \theta, \gamma_1, \gamma_2)&=&(N+1)^{-1} \Biggl\{ \gamma_1 \ln \left( e^{e^{-\left(\frac{t_{1_{1:N}}}{\theta}\right)^{-\lambda}}}-1 \right) \nonumber \\
    && + \sum_{i=2}^{N} \ln \left[ \left( e^{e^{-\left(\frac{t_{1_{i:N}}}{\theta}\right)^{-\lambda}}}-1 \right)^{\gamma_1} - \left( e^{e^{-\left(\frac{t_{1_{i-1:N}}}{\theta}\right)^{-\lambda}}}-1 \right)^{\gamma_1} \right]  \nonumber\\
    && + \ln\left[ (e-1)^{\gamma_1} - \left( e^{e^{-\left(\frac{t_{1_{N:N}}}{\theta}\right)^{-\lambda}}}-1 \right)^{\gamma_1} \right] - (N+1)\, \gamma_1 \,\ln(e-1)\Biggl\} \nonumber \\
    && +(M+1)^{-1} \Biggl\{ \gamma_2 \ln \left( e^{e^{-\left(\frac{t_{2_{1:M}}}{\theta}\right)^{-\lambda}}}-1 \right)  \label{eq:5.7} \\
    && + \sum_{j=2}^{M} \ln \left[ \left( e^{e^{-\left(\frac{t_{2_{j:M}}}{\theta}\right)^{-\lambda}}}-1 \right)^{\gamma_2} - \left( e^{e^{-\left(\frac{t_{2_{j-1:M}}}{\theta}\right)^{-\lambda}}}-1 \right)^{\gamma_2} \right] \nonumber \\
    && + \ln\left[ (e-1)^{\gamma_2} - \left( e^{e^{-\left(\frac{t_{2_{M:M}}}{\theta}\right)^{-\lambda}}}-1 \right)^{\gamma_2} \right] - (M+1)\, \gamma_2 \,\ln(e-1)\Biggl\}. \nonumber
\end{eqnarray}
Partial differentiation of \eqref{eq:5.7} with respect to each of the parameters and equating them to zero will yield the corresponding MPSEs, but cannot be solved explicitly as the case of MLEs.

Solving $\frac{\partial \, \ln PS}{\partial\, \gamma_1} = 0$ and $\frac{\partial \, \ln PS}{\partial\, \gamma_2} = 0$ yield the estimators $\hat{\gamma_1}^{MPSE}$ and $\hat{\gamma_2}^{MPSE}$ of $\gamma_1$ and $\gamma_2$ respectively. Hence,
\begin{equation}
    \hat{R}^{MPSE}=\frac{\hat{\gamma_1}^{MPSE}}{\hat{\gamma_1}^{MPSE} + \hat{\gamma_2}^{MPSE}}.
    \label{eq:5.8}
\end{equation}

\subsection{Multi-component, $R_{c,k}$}
Consider $k$ components, $T_{1}^{1}, T_{1}^{2}, ..., T_{1}^{k} \sim PGDUS-IW (\lambda, \theta, \gamma_1)$ distribution and $T_2 \sim PGDUS-IW (\lambda, \theta, \gamma_2)$ distribution.
Then, the multi-component system reliability $R_{c,k}$ is
\begin{eqnarray}
  R_{c,k} &=& Prob(atleast\, c \,of\, the\, T_{1}^{l}s\, exceed\, T_2)\, ; l = 1, 2, ..., k \nonumber \\
  &=& \sum_{l=c}^{k} \binom{k}{l} \int_{t=0}^{\infty}[1-F_{T_{1_{PGDUS-IW}}}(t)]^{l}[F_{T_{1_{PGDUS-IW}}}(t)]^{k-l} dF_{T_{2_{PGDUS-IW}}}(t)  \nonumber
\end{eqnarray}
On proper substitution and simplification, we get
\begin{eqnarray}
     R_{c,k} &=& \sum_{l=c}^{k}\binom{k}{l} \frac{\gamma_2 \lambda \theta^\lambda}{(e-1)^{\gamma_2+k\gamma_1}} \int_{0}^{\infty} t^{-(\lambda+1)} \, e^{-\left(\frac{t}{\theta}\right)^{-\lambda}}\, e^{e^{-\left(\frac{t}{\theta}\right)^{-\lambda}}} \, \left[e^{e^{-\left(\frac{t}{\theta}\right)^{-\lambda}}}-1\right]^{\gamma_1(k-l)+\gamma_2-1}\nonumber \\
     &&\left[(e-1)^{\gamma_1} - \left(e^{e^{-\left(\frac{t}{\theta}\right)^{-\lambda}}}-1\right)^{\gamma_1} \right]^{l}   \cdot dt \nonumber
\end{eqnarray}
and on solving the integral, the multi-component parameter is obtained as
\begin{equation}
    R_{c,k}= \sum_{l=c}^{k} \,\sum_{p=0}^{l}\, \binom{k}{l}\binom{l}{p} \frac{(-1)^p \gamma_2}{\gamma_1(k+p-l)+\gamma_2}; \hspace{2mm} \gamma_1, \gamma_2 >0.
    \label{eq:5.9}
\end{equation}

\subsubsection{ML Estimation of $R_{c,k}$}
We take random samples of size $N$, $T_{1_1}^{l}, T_{1_2}^{l}, ..., T_{1_N}^{l}$, from $PGDUS-IW(\lambda, \theta, \gamma_1)$, where $l=1, 2, ..., k$ and $T_{2_1}, T_{2_2}, ..., T_{2_N}$, from $PGDUS-IW(\lambda, \theta, \gamma_2)$ distribution. To estimate $R_{c,k}$, we find the estimators of $\gamma_1$ and $\gamma_2$. The likelihood function is
\begin{equation*}
    L(\lambda, \theta, \gamma_1, \gamma_2 \mid t_{1_{j}}^{l}, t_{2_j})=\prod_{j=1}^{N} \left\{ \left[\prod_{l=1}^{k}  f_{T_{1_{PGDUS}}}(t_{1_{j}}^{l}, \lambda, \theta, \gamma_1)\right] f_{T_{2_{PGDUS}}}(t_{2_j}, \lambda, \theta, \gamma_2) \right\}.
\end{equation*}
\noindent On required substitution, simplification and taking logarithm, we get
\begin{eqnarray}
    \ln L&=& N k\ln \gamma_1 + N \ln \gamma_2 + N(k+1) (\ln \lambda + \lambda \ln \theta) - N(k \gamma_1 +  \gamma_2) \ln (e-1) \nonumber \\
    && - (\lambda+1) \left[ \sum_{j=1}^{N}\sum_{l=1}^{k} \ln t_{1_{j}}^{l} + \sum_{j=1}^{N} \ln t_{2_j}\right] - \sum_{j=1}^{N}\sum_{l=1}^{k} \left(\frac{t_{1_{j}}^{l}}{\theta}\right)^{-\lambda}- \sum_{j=1}^{N} \left(\frac{t_{2_j}}{\theta}\right)^{-\lambda} + \sum_{j=1}^{N}\sum_{l=1}^{k}  e^{-\left(\frac{t_{1_{j}}^{l}}{\theta}\right)^{-\lambda}} \nonumber \\
    &&  + \sum_{j=1}^{N} e^{-\left(\frac{t_{2_j}}{\theta}\right)^{-\lambda}} + (\gamma_1-1)\sum_{j=1}^{N}\sum_{l=1}^{k} \ln \left(e^{e^{-\left(\frac{t_{1_{j}}^{l}}{\theta}\right)^{-\lambda}}} -1\right) + (\gamma_2-1)\sum_{j=1}^{N} \ln \left(e^{e^{-\left(\frac{t_{2_j}}{\theta}\right)^{-\lambda}}} -1\right). \nonumber \\
    \label{eq:5.10}
\end{eqnarray}
On partial differentiation of \eqref{eq:5.10} and on solving we get,
\begin{equation}
    \hat{\gamma_1}^{MLE}=\frac{Nk}{Nk \ln (e-1)-\sum_{j=1}^{N}\sum_{l=1}^{k} \ln \left(e^{e^{-\left(\frac{t_{1_{j}}^{l}}{\theta}\right)^{-\lambda}}} -1\right)},
    \label{eq:5.11}
\end{equation}
which is the MLE of $\gamma_1$ and
\begin{equation}
    \hat{\gamma_2}^{MLE}=\frac{N}{N \ln (e-1)-\sum_{j=1}^{N}\ln \left(e^{e^{-\left(\frac{t_{2_j}}{\theta}\right)^{-\lambda}}} -1\right)},
    \label{eq:5.12}
\end{equation}
which is the MLE of $\gamma_2$. Hence, the MLE of $R_{c,k}$ is
\begin{equation}
    \hat{R_{c,k}}^{MLE}=\sum_{l=c}^{k} \,\sum_{p=0}^{l}\, \binom{k}{l}\binom{l}{p} \frac{(-1)^p \hat{\gamma_2}^{MLE}}{\hat{\gamma_1}^{MLE}(k+p-l)+\hat{\gamma_2}^{MLE}}.
    \label{eq:5.13}
\end{equation}

\subsubsection{MPS Estimation of $R_{c,k}$}
The ordered random samples $T_{1_{1:N}}^{l} < T_{1_{2:N}}^{l} < . . . < T_{1_{N:N}}^{l}$ taken from $PGDUS-IW (\lambda, \theta, \gamma_1)$ distribution, where $l=1, 2, ..., k$ and another one $T_{2_{1:N}} < T_{2_{2:N}} < . . . < T_{2_{N:N}}$ from $PGDUS-IW (\lambda, \theta, \gamma_2)$ distribution. i.e., $Nk$ observations are taken from $T_1$'s distribution ($T_{1_{1:Nk}} < T_{1_{2:Nk}} < . . . < T_{1_{Nk:Nk}}$) and $N$ observations are taken from $T_2$'s distribution.

Following the similar computations carried out in single component parameter $R$, here also we need to maximize
\begin{equation*}
    \ln PS(\lambda, \theta, \gamma_1, \gamma_2) = \ln PS_1(\lambda, \theta, \gamma_1) + \ln PS_2(\lambda, \theta, \gamma_2),
\end{equation*}
where, $PS_1$ is the PS function of $T_1$'s distribution and $PS_2$ is the PS function of $T_2$'s distribution.

\noindent Here, we obtain
\begin{eqnarray}
    \ln PS(\lambda, \theta, \gamma_1, \gamma_2)&=&(Nk+1)^{-1} \Biggl\{ \gamma_1 \ln \left( e^{e^{-\left(\frac{t_{1_{1:Nk}}}{\theta}\right)^{-\lambda}}}-1 \right) \nonumber \\
    && + \sum_{i=2}^{Nk} \ln \left[ \left( e^{e^{-\left(\frac{t_{1_{i:Nk}}}{\theta}\right)^{-\lambda}}}-1 \right)^{\gamma_1} - \left( e^{e^{-\left(\frac{t_{1_{i-1:Nk}}}{\theta}\right)^{-\lambda}}}-1 \right)^{\gamma_1} \right]  \nonumber\\
    && + \ln\left[ (e-1)^{\gamma_1} - \left( e^{e^{-\left(\frac{t_{1_{Nk:Nk}}}{\theta}\right)^{-\lambda}}}-1 \right)^{\gamma_1} \right] - (Nk+1)\, \gamma_1 \,\ln(e-1)\Biggl\} \nonumber \\
    && +(N+1)^{-1} \Biggl\{ \gamma_2 \ln \left( e^{e^{-\left(\frac{t_{2_{1:N}}}{\theta}\right)^{-\lambda}}}-1 \right)  \label{eq:5.14} \\
    && + \sum_{j=2}^{N} \ln \left[ \left( e^{e^{-\left(\frac{t_{2_{j:N}}}{\theta}\right)^{-\lambda}}}-1 \right)^{\gamma_2} - \left( e^{e^{-\left(\frac{t_{2_{j-1:N}}}{\theta}\right)^{-\lambda}}}-1 \right)^{\gamma_2} \right] \nonumber \\
    && + \ln\left[ (e-1)^{\gamma_2} - \left( e^{e^{-\left(\frac{t_{2_{N:N}}}{\theta}\right)^{-\lambda}}}-1 \right)^{\gamma_2} \right] - (N+1)\, \gamma_2 \,\ln(e-1)\Biggl\}. \nonumber
\end{eqnarray}
Partial differentiation of \eqref{eq:5.14} with respect to each of the parameters and equating them to zero will yield the corresponding MPSEs, but here also it cannot be solved explicitly.

Solving $\frac{\partial \, \ln PS}{\partial\, \gamma_1} = 0$ and $\frac{\partial \, \ln PS}{\partial\, \gamma_2} = 0$ yield the estimators $\hat{\gamma_1}^{MPSE}$ and $\hat{\gamma_2}^{MPSE}$ of $\gamma_1$ and $\gamma_2$ respectively. Hence,
\begin{equation}
    \hat{R_{c,k}}^{MPSE}=\sum_{l=c}^{k} \,\sum_{p=0}^{l}\, \binom{k}{l}\binom{l}{p} \frac{(-1)^p \hat{\gamma_2}^{MPSE}}{\hat{\gamma_1}^{MPSE}(k+p-l)+\hat{\gamma_2}^{MPSE}}.
    \label{eq:5.15}
\end{equation}

\section{\textbf{CONCLUSION}}
\label{Sec6}
The main goal of this study was to present and thoroughly examine the PGDUS-IW model, a unique lifetime distribution that has important applications in reliability theory, especially for parallel systems. We determined and investigated the PGDUS-IW model's basic statistical characteristics, such as moments, MGF, CF, CGF, entropy, extropy, and more. We used ML and MPS estimation techniques to estimate the model parameters, and we provided comprehensive steps and a simulation study to verify the estimation process. The effectiveness of the novel model in fitting a variety of real-life datasets is demonstrated through a study of real-world data that compares it to the previously proposed PGDUS models with three parameters. Additionally, we estimated these parameters using the same estimation methodologies and corresponding simulation studies after deriving $P(T_2<T_1)$ in both single and multi-component models. The results of our examination of real-world data were used to fit the single component model and compare the estimating methods used to fit the two datasets. Having been considered, this thorough investigation demonstrates its resilience and suitability for use in reliability theory, especially for parallel systems, offering a useful resource for researchers working in this field.

\end{document}